\documentclass[aps,amsmath,amssymb,superscriptaddress,prb,10pt,twocolumn,eqrefnum]{revtex4-2}
\usepackage{amsmath}    
\usepackage{graphicx}  
\usepackage{verbatim}  
\usepackage{color}      
\usepackage{subfigure} 
\usepackage{hyperref}   
\usepackage{xcolor,txfonts,mathtools,mathrsfs}
\usepackage{bm}

\definecolor{SC-color}{named}{blue}

\definecolor{SCh-color}{named}{green}

\definecolor{Note-color}{named}{red}

\definecolor{SHH-color}{named}{magenta}

\newcommand{\beq}{\begin{equation}}
\newcommand{\eeq}{\end{equation}}
\newcommand{\Beq}{\begin{eqnarray}}
\newcommand{\Eeq}{\end{eqnarray}}
\newcommand{\bml}{\begin{multline}}

\newcommand{\bea}{\begin{align}}
\newcommand{\ena}{\end{align}}
\newcommand{\bsp}{\begin{split}}
\newcommand{\esp}{\end{split}}

\begin{document}
\title{Microscopic theory of supercurrent suppression by gate-controlled surface depairing }
\author{Subrata Chakraborty}
\email[Correspondence to: ]{subrata.chakraborty@uni-konstanz.de}
\affiliation{Fachbereich Physik, Universit{\"a}t Konstanz, D-78457 Konstanz, Germany}
\author{Danilo Nikoli\'c}
\email{danilo.nikolic@uni-konstanz.de}
\affiliation{Fachbereich Physik, Universit{\"a}t Konstanz, D-78457 Konstanz, Germany}
\affiliation{Institut f\"ur Physik, Universit\"at Greifswald, Felix-Hausdorff-Strasse 6, 17489 Greifswald, Germany}
\author{Juan Carlos Cuevas}
%\email{juancarlos.cuevas@uam.es}
\affiliation{Departamento de F{\'i}sica Te{\'o}rica de la Materia Condensada and Condensed Matter Physics Center (IFIMAC),
Universidad Aut{\'o}noma de Madrid, E-28049 Madrid, Spain}
\author{Francesco Giazotto}
%\email{francesco.giazotto@sns.it}
\affiliation{NEST, Istituto Nanoscienze-CNR and Scuola Normale Superiore, Pisa I-56127, Italy}
\author{Angelo Di Bernardo}
%\email{angelo.dibernardo@uni-konstanz.de}
\affiliation{Fachbereich Physik, Universit{\"a}t Konstanz, D-78457 Konstanz, Germany}
\author{Elke Scheer}
%\email{elke.scheer@uni-konstanz.de}
\affiliation{Fachbereich Physik, Universit{\"a}t Konstanz, D-78457 Konstanz, Germany}
\author{Mario Cuoco}
%\email{mario.cuoco@spin.cnr.it}
\affiliation{SPIN-CNR, c/o Universit{\`a} degli Studi di Salerno, I-84084 Fisciano (Salerno), Italy}
\author{Wolfgang Belzig}
%\email{wolfgang.belzig@uni-konstanz.de}
\affiliation{Fachbereich Physik, Universit{\"a}t Konstanz, D-78457 Konstanz, Germany}
\date{\today}

\begin{abstract} 
Recently gate-mediated supercurrent suppression in superconducting nano-bridges has been reported in many experiments. This could be either a direct or an indirect gate effect. The microscopic understanding of this observation is not clear till now. Using the quasiclassical Green's function method, we show that a small concentration of magnetic impurities at the surface of the bridges can significantly help to suppress superconductivity and hence the supercurrent inside the systems while applying a gate field. This is because the gate field can enhance the depairing through the exchange interaction between the magnetic impurities at the surface and the superconductor. We also obtain a \emph{symmetric} suppression of the supercurrent with respect to the gate field, a signature of a direct gate effect. We discuss the parameters range of applicability of our 
model and how it is able to qualitatively capture the main aspects of 
the experimental observations. Future experiments can verify our predictions by modifying the surface with magnetic impurities. 

\end{abstract}
\maketitle

{\section{Introduction}}\label{Sec:Intro}

The role of an external magnetic field in superconductors has been thoroughly analyzed in the past. In contrast, the investigation of electric-field-mediated physics in such systems was not popular till the last decade. Usually, the electric field's effect in a bulk superconducting system is insignificant due to its small penetration depth \cite{ashcroft, tinkham, Choi2014, Piatti2016, Piatti2017, Ummarino2017}. However, the role of an electric field in a thin-film superconductor can be significant enough. Recent experiments demonstrate that a large electric field from a gate can control the supercurrent in a superconducting nano-bridge \cite{Francesco2018, Francesco2020, Francesco2021, Golokolenov2021, Alegria2021, Ritter2022, Aprili2021, Moodera2021}. Namely, at low temperatures, the supercurrent flowing along the bridge monotonically decays by  increasing the gate field. In addition, it has been found that the critical electric field, at which the supercurrent vanishes, is robust with respect to experimental temperatures \cite{Francesco2018, Francesco2020, Francesco2021}. It has been also found that the critical gate field is marginally affected by a weak external magnetic field applied across the bridge \cite{Francesco2018,Boursprr}.   

From several experimental evidences, many distinct physical mechanisms have been proposed to describe this effect that we call gate-controlled supercurrent (GCS) suppression \cite{supergatereview}, a set of which suggests that the gate field could cause Cooper pair breaking resulting in direct supercurrent suppression \cite{Francesco2018, Francesco2020, Francesco2021}. On the other hand, several experimental reports disagree with the suppression mechanism due to the direct field coupling to the bridge by the observation of a non-vanishing leakage current between the gate and the bridge for large gate fields \cite{Golokolenov2021, Alegria2021, Ritter2022, Aprili2021}. 
%%%%%
 \begin{figure}[t!]
\includegraphics[width=0.96\linewidth]{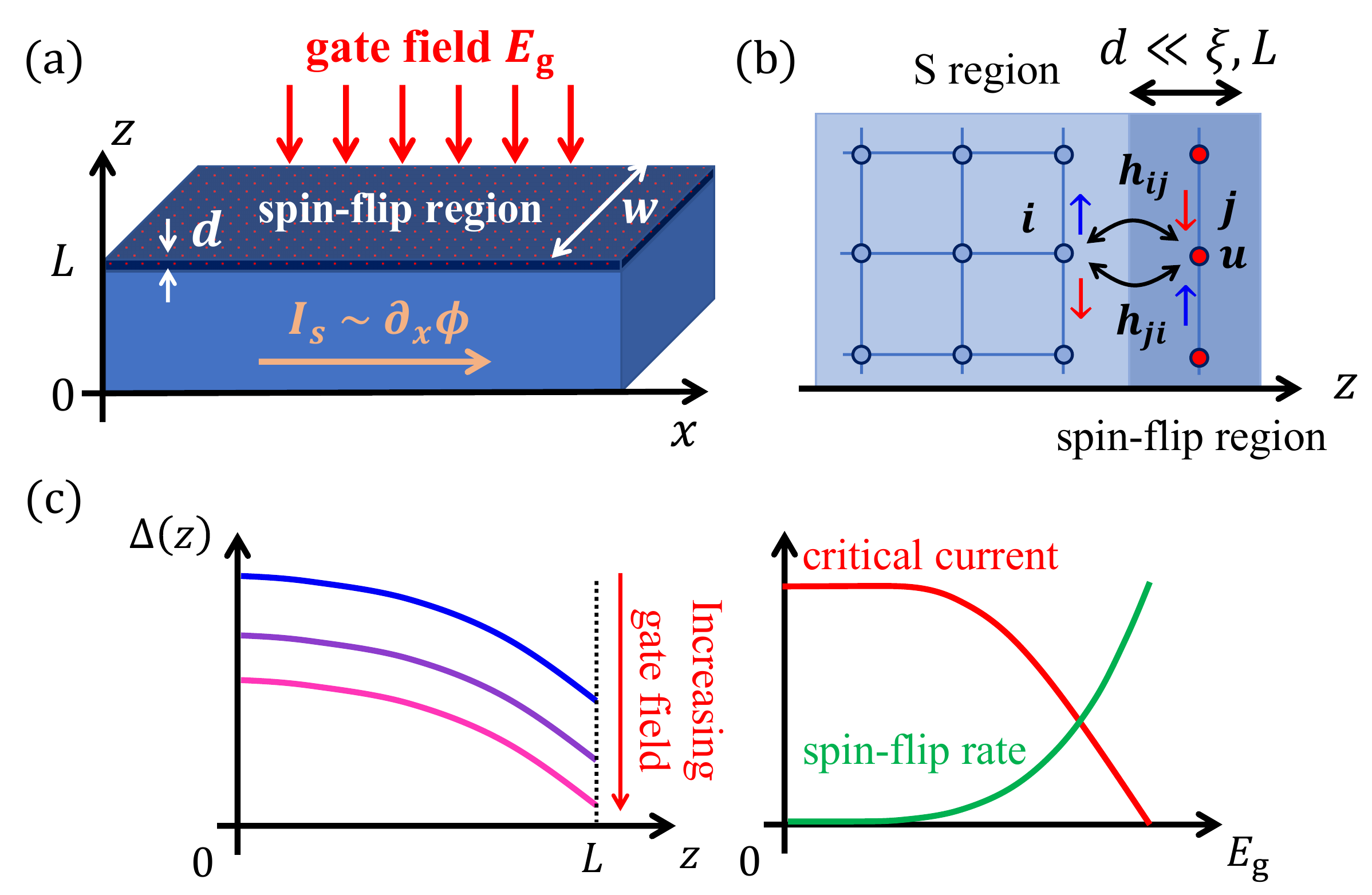}
\caption{\label{schematic}{ (a) Scheme of a superconducting bridge of width $w$, thickness $L$, and infinite length. A finite supercurrent, $I_s$, flows along the bridge ($x$-direction) due to a constant phase gradient, $\partial_x\phi$. A gate electric field $E_g$ is applied across the bridge ($z$-direction). Due to the Thomas-Fermi-like screening, the field effectively penetrates the bridge's surface up to length $d\ll L,\xi$ 
enhancing spin-flip scattering at the surface magnetic impurity centers ($\xi$ is the superconducting coherence length). (b) The gate-mediated electron hopping between a superconducting site $i$ and a magnetic impurity site $j$ in the surface layer is $h_{ij}(E_{g})$, which causes spin-flip scattering in the surface layer. Electron pair repulsion at a magnetic impurity site is $u$. (c) (Left panel) Schematic representation of the superconducting gap profile, $\Delta(z)$, closing across the bridge for increasing values of the gate field and (right panel) the critical supercurrent suppression due to gate-dependent spin-flip scattering. }}
\end{figure} 
%%%%%%
The leakage current can cause supercurrent suppression either via high-energy quasiparticle injection to the bridge \cite{Golokolenov2021, Alegria2021, Ritter2022}, or indirectly via phonon-induced Joule heating of the system due to charge injection into the substrate \cite{Ritter2021a, Catto}. The leakage current can also cause phase fluctuations in the bridge resulting in an indirect suppression by generating a nonequilibrium phonon state at the substrate and in the superconductor~\cite{Ritter2022,Aprili2021, Elalaily2022}. However, the leakage-current-mediated indirect mechanisms can be restricted by suitable experimental setups \cite{Francesco2020, Francesco2021, supergatereview}. Here, our objective is to have a theoretical description of the direct field effect.  
 
Earlier microscopic theoretical analyses of the direct field effect were based on electric field-induced surface orbital polarization \cite{SOP2020}, Rashba-like surface effects \cite{Boursprr, Rashba2021, Mercaldoprr, Chirolliprl}, and excitation of a superconducting state because of Schwinger-like effects \cite{Schwinger2021}. Recent theoretical works based on the Ginzburg-Landau paradigm also analyze a phenomenological direct field effect \cite{Amoretti_GL} or predict a spin-orbit enhanced surface barrier in combination with a magnetic insulator \cite{Akashprb}. However, none of these analyses fully describes the GCS effect. A fully microscopic theory, which accounts for the experimental fact, is necessary for the complete understanding of the GCS suppression.

In this work, we investigate the gate-field-mediated direct supercurrent suppression in a long superconducting  nano-bridge [see Fig.~\ref{schematic}(a)]. A uniform supercurrent flows along the bridge and the gate field is applied across the bridge. We consider the presence of a small concentration of magnetic impurities in the surface layer of the bridge [see Fig.~\ref{schematic}(b)]. It is known that oxide layer forming on the surface of some superconductors upon exposure to air can host paramagnetic impurities \cite{Sheridan21, Graaf17, Adelstein17, Kharitonov12, Thomas11}. These magnetic impurities can introduce decoherence channels \cite{Sheridan21} or increase flux noise \cite{Wang2015}, thus affecting the performance of devices like superconducting resonators \cite{Graaf18} or of quantum circuits \cite{Wang2015}. 
Our model is based on the assumption that a large gate field can significantly enhance  surface depairing via spin-flip scattering processes at the magnetic impurity centers [see Fig.~\ref{schematic}(b)].
Using the quasiclassical Green's function (GF) formalism, we find that a sufficiently strong gate-induced surface depairing can cause superconducting gap quenching across the bridge [see Fig.~\ref{schematic}(c)]. Consequently, the critical current decreases by increasing the gate field [see Fig.~\ref{schematic}(c)]. We also analyze the impact of finite temperatures and weak magnetic fields across the bridge. Finally, we relate our model with real experimental parameters.
 
The rest of the paper is organized as follows. First, in Sec.~\ref{Sec:Model} we present our theoretical model based on microscopic quasiclassical theory to describe GCS suppression. Then, in Sec.~\ref{Sec:Results} we discuss our main results on the superconducting gap quenching and supercurrent suppression by the gate-mediated surface depairing. Finaly, we summarize the main conclusions of 
this work in Sec.~\ref{Sec:Conclusions}.Furthermore, in Appendix~\ref{Appendix:GL} we present an effective Ginzburg-Landau model based on our full microscopic theory, and in Appendix~\ref{Appendix:Phenomenology} we discuss the external electric field-mediated critical current suppression by phenomenologically introducing temperature- and magnetic field-dependent relative permittivity.

\vspace{5mm}
\section{Model}\label{Sec:Model}

\subsection{System}
We begin with an infinitely long superconducting nano-bridge of thickness $L$, depicted in Fig.~\ref{schematic}(a). We assume a uniform supercurrent, $I_s$, flowing along the bridge ($x$-direction) due to a constant superconducting phase gradient $\partial_x\phi$ with the phase $\phi({\bf{r}})$. The gate field, $E_g$, is applied across the bridge ($z$-direction) from a gate electrode (not shown in the schematic). Note that the $z$-coordinate runs from $0$ to $L$ inside the bridge. Due to the Thomas-Fermi screening, the gate field effectively penetrates the superconductor over a rather small length $d\ll L,~\xi$, where $\xi$ is the superconducting coherence length~\cite{ashcroft}. We assume the presence of a small concentration of magnetic impurities in the surface layer of thickness $d$. The main assumption in our model is that a large $E_g$ causes a significantly high spin-exchange coupling between the magnetic impurity sites and the superconducting sites through gate-induced electron hopping, $h_{ij}(E_g)$ [see Fig.~\ref{schematic}(b)]. The hopping is linear in the gate field, $h_{ij}(E_g)\sim E_g$, in the lowest order of the perturbation theory, and this can result in a strong surface spin-flip scattering leading to the GCS suppression. 
Note that in this work we do not consider the presence of magnetic impurities inside the bridge as they can hardly experience the gate field, although this scenario would not affect our final conclusions qualitatively.

To describe the supercurrent, we use the quasiclassical GF formalism in equilibrium \cite{LarkinOvchinnikov, Belzig1999}. We assume the diffusive limit, elastic mean free path $\ll \xi=\sqrt{\hbar D/2k_BT_c}$ where $D$ is the diffusion coefficient of the material and $T_c$ is the bulk superconducting critical temperature. In the absence of spin-splitting, we can work in Nambu space, where the GF matrix can be parametrized as $\hat{g}=G\hat{\tau}_3 +F\hat{\tau}_+ +F^\dag\hat{\tau}_-$  and is subject to the normalization $\hat{g}^2=\hat{\tau}_0\implies G^2+FF^\dagger=1$. Here, $\hat{\tau}_{\pm}=(\hat{\tau}_1 \pm i\hat{\tau}_2)/2$ and $\hat{\tau}_i$ are the Pauli matrices in Nambu space. The GF matrix in the bridge satisfies the Usadel equation~\cite{Usadel} 
\Beq
&& \hbar D\bm{\nabla}\cdot(\hat{g}\bm{\nabla}\hat{g})=[\omega_n\hat{\tau}_3 +\Delta ({\bf{r}})\hat{\tau}_+
+\Delta^*({\bf{r}})\hat{\tau}_- +\hat{\Sigma},\hat{g}], \label{eq2}
\Eeq
where $\omega_n=(2n +1)\pi k_BT$ defines Matsubara frequencies at temperature $T$ with $n=0, \pm 1, \pm 2,\dots$ and $\Delta(\bf{r})$ is the inhomogeneous superconducting order parameter. In addition, we account for the depairing effects described by the self-energy $\hat{\Sigma}=[3\Gamma_\mathrm{sf}+\Gamma_\mathrm{orb}(B)]\hat{\tau}_3\hat{g}\hat{\tau}_3$. 
Here, $\Gamma_\mathrm{sf}$ is the spin-flip scattering rate and $\Gamma_\mathrm{orb}(B)=(\Delta_0/4)(B/B_c)^2$
is the orbital depairing rate due to a weak external magnetic field $B$ across the bridge [$z$-direction; see Fig.~\ref{schematic}(a)]. In the latter term, $B_c=\sqrt{3\Delta_0/(\hbar D)}(\Phi_0/ \pi w)$ is the critical magnetic field of a bare BCS superconducting bridge of width $w$ with $\Phi_0=h/(2e)$ being the magnetic flux quantum \cite{Belzig1996}. The BCS gap at zero temperature is $\Delta_0=1.764k_BT_c$.
As discussed below, the spin-flip scattering is present only in a thin surface layer and effectively manifests itself as a boundary condition (see Sec.~\ref{Sec:BC}). We consider that $L$ and $w$ are smaller than the London penetration depth.

Due to a constant supercurrent along $x$-direction, we account for the spatial inhomogeneity of the system by introducing the ansatz: $\Delta({\bf{r}})=\Delta(z)e^{i\phi(x)}$, $F({\bf{r}},\omega_n)=f(z,\omega_n)e^{i\phi(x)}$ and $G({\bf{r}},\omega_n)=g(z,\omega_n)$ with superconducting phase $\phi(x)=qx$. Hence, the phase gradient along the bridge is $\partial_x\phi=q$ providing a uniform current. We utilize the so-called $\theta$-parametrization, $g(z,\omega_n)=\cos\theta(z,\omega_n)$ and $f(z,\omega_n)=\sin\theta(z,\omega_n)$ \cite{Belzig1999, Nazarov}, obtaining
\Beq
\hbar D\partial^2_z\theta + 2\Delta(z)\cos\theta -2\omega_n\sin\theta -2\Gamma_\mathrm{eff}(q)\sin \theta\cos\theta =0, ~  \label{eq3}
\Eeq
where $\Gamma_\mathrm{eff}(q)=\hbar D(q^2/2) +2\Gamma_{\rm{orb}}+6 \Gamma_\mathrm{sf}$ is an effective pair-breaking rate. 
Note that $\Gamma_{\rm{sf}}=0$ inside the bridge due to the lack of magnetic impurities, which are, as already mentioned, present only at the surface in our model.

To obtain a full solution to the problem, the superconducting gap across the bridge should be treated self-consistently as
\beq
\Delta(z)\ln(T/T_c)=2\pi k_BT\sum_{n=0}^{N_D(T)}\left[ \sin\theta(z,\omega_n) -\frac{\Delta(z)}{\omega_n} \right], \label{eq8}
\eeq
where $N_D$ truncates the summation over $\omega_n$ up to the Debye frequency. Based on the preceding discussion, we can calculate the supercurrent density along the bridge as 
\Beq
J_s(y,z)  &=&\frac{2\sigma_N}{e}\pi k_BTq\sum_{n=0}^{N_D(T)} \sin^2\theta(z,\omega_n), \label{eq9}
\Eeq
where $\sigma_N=2e^2\nu D$ is the normal-state conductivity. The supercurrent itself is calculated by integrating the expression above over the cross-section of the bridge, i.e., $I_s(q)=\int dy~dz~J_s(y,z;q)$, and the critical supercurrent is obtained as $I_c=I_s(q=q_{\rm{max}})$, such that $I_s$ is maximum at $q=q_{\rm{max}}$. Here, we express the supercurrent in the units of $I_{\rm{sc}}=2\pi \sigma_N Lwk_BT_c/(e\xi)$. 

As already anticipated, obtaining a full solution of Eq.~\eqref{eq3} requires us to apply appropriate boundary conditions and here $E_g$-dependent spin-flip processes enter playing a crucial role. In what follows, we first discuss the effect of an external gate field on the spin-flip scattering rate and then how it effectively translates into the pair-breaking boundary conditions.

\subsection{Gate-induced surface spin-flip scattering}\label{Sec:BC}

Here we demonstrate how the gate field $E_g$ can participate in magnetic impurity scattering in the surface layer. The finite gate field penetrating a thin surface layer of thickness $d$ along the $z$ direction can be expressed as $E_g=-\partial V_g/\partial z$. Considering a uniform electric field near $z=L$, the scalar potential within the surface layer becomes $V_g\approx -E_gz$. Hence, the gate field $E_g$ can drive the electron hopping process between the superconducting sites and the magnetic impurity centers in this region. Gate field can also induce electron hopping between the superconducting sites and the magnetic impurity centers in the surface layer via spin-orbit interactions \cite{Ishibashi_2018, Ando_2018, Mankovsky}. For simplicity, we here only express electron hopping due to the scalar potential $V_g$ as 
\Beq
t_{ij}(E_g)=-eE_g\int d^3r~\psi^*_s({\bf{r,R}}_i)~z~ \psi_{m}({\bf{r,R}}_j), \label{eq10}
\Eeq
where ${\bf{r}}$ and $z$ stand for one electron coordinate and $\psi_{s/m}({\bf{r}},{\bf{R}}_{i/j})$ stands for the electronic wave function at a superconductor/ magnetic impurity center located at spatial coordinate ${\bf{R}}_{i/j}$. 
 
The gate-modulated electron hopping between the superconducting and the magnetic impurity sites leads to electron spin-exchange processes resulting in an effective spin-exchange Hamiltonian
\Beq
\hat{H}_{ij}^{\rm{ex}} &=& \frac{\left[t^{(0)}_{ij}+t_{ij}(E_g)\right]^2}{2u}{\text{\boldmath{$\sigma$}}}(i)\cdot {\bf{s}}(j) \nonumber \\
&=& \frac{h_{ij}^2(E_g)}{2u} {\text{\boldmath{$\sigma$}}}(i)\cdot {\bf{s}}(j)
=J_{ij}(E_g) {\text{\boldmath{$\sigma$}}}(i)\cdot {\bf{s}}(j), \label{eq11}
\Eeq
where $t_{ij}^{(0)}$ accounts for the corresponding hopping process in the absence of $E_g$, $h_{ij}=t_{ij}^{(0)} + t_{ij}(E_g)$ [see Fig.~\ref{schematic}(b)], $u$ stands for the electron-pairing energy at a magnetic impurity site [see Fig.~\ref{schematic}(b)], and ${\text{\boldmath{$\sigma$}}}(i)$ and ${\bf{s}}(j)$ stand for the Pauli spin matrices at the superconducting site ${\bf{R}}_i$ and the magnetic impurity site ${\bf{R}}_j$, respectively. Apparently, the exchange energy $J_{ij}(E_g)$ in Eq.~\eqref{eq11} modulates with $E_g$. This is similar to the recent studies about the impact of an external electric field on the electron spin-exchange interaction \cite{Ishibashi_2018, Ando_2018, Mankovsky}. 

Due to the spin-exchange mechanisms described above, superconducting electrons can scatter with the magnetic impurity centers. Considering magnetic moments of the magnetic impurity centers as classical spins, the corresponding spin-flip self-energy that enters the Usadel equation~\eqref{eq2} in the surface layer reads~\cite{kopnin}
\Beq 
&& \hat{\Sigma}_{\rm{sf}}(E_g)=3\Gamma_{\rm{sf}}(E_g)~\hat{\tau}_3 \hat{g} \hat{\tau}_3,~~ \label{jan10a} 
\Eeq
where a numerical factor of 3 is due to the summation over the spin degrees of freedom. Defined in this way, the spin-flip self-energy enters Eq.~\eqref{eq2}. The spin-flip scattering rate itself is given by 
\Beq
 \Gamma_{\rm{sf}}(E_g)=\frac{2\pi}{3} \nu N_m {\left|\langle{\bf{s}}\rangle\right|^2} \int \frac{d\Omega}{4\pi} \left|J(\Theta,E_g)\right|^2, \label{eq13} 
\Eeq
where $\nu$ is the density of states at the Fermi level, $N_m$ is the density of the magnetic impurities in the surface layer, $\left|\langle{\bf{s}}\rangle\right|$ defines the magnitude of the average classical magnetic moment of a magnetic impurity center.
In addition, $J(\Theta,E_g)$ is the Fourier transform of $J_{ij}(E_g)$ under quasiclassical scheme:
\Beq
&& J_{ij}(E_g)=\int d^3p~e^{i{\bf{p}}\cdot({\bf{r}}_i - {\bf{r}}_j)} J({\bf{p}},E_g), \label{eq14} \\
&&  \left. J({\bf{p}}-{\bf{p^\prime}},E_g)\right|_{p=p^\prime =p_F} = J(\Theta,E_g), \label{eq15}
\Eeq
where $p_F$ is the magnitude of the Fermi momentum.
Considering Eqs.~\eqref{eq10}-\eqref{eq15}, we can in general express the spin-flip scattering rate with respect to $E_g$ as $\Gamma_{\rm{sf}}(E_g)=\sum_{i=0}^4A_iE^i_g$. 

Due to the random spatial distribution of the magnetic impurity centers, hopping amplitudes $t_{ij}^{(0)}$ are random in real space. For perfectly random distribution of $t^{(0)}_{ij}$, upon the Fourier transformation above we can express the spin-flip scattering rate as $\Gamma_{\rm{sf}}(E_g)=A_0+A_2E^2_g+A_4E^4_g$. 
The coefficients $A_0 \sim \left(t^{(0)}_{ij}\right)^4$ and $A_2\sim \left(t^{(0)}_{ij}\right)^2$ would increase with the increasing strength of $t^{(0)}_{ij}$. For large $E_g$, with relatively insignificant impact of $t^{(0)}_{ij}$, we may consider $\Gamma_{\rm{sf}}(E_g)\sim  E^4_g>0$ for large gate fields. The even behavior of $\Gamma_{\rm{sf}}$ vs  $E_g$ yields bipolarity in gating as observed in experiments \cite{Francesco2018, Francesco2020, Francesco2021,Boursprr}, a very relevant fingerprint of the above proposed mechanism.

{\subsection{Boundary conditions}}

This gate-mediated spin-flip scattering of superconducting electrons on the magnetic impurities sitting at the surface leads to the Cooper pair breaking. Now, we show how the described mechanism results in a gate-dependent boundary condition. By assuming that the large $E_g$ makes the spin-flip processes energetically dominant, the Usadel equation [see Eq.~\eqref{eq2}] close to the surface adopts the form
\Beq
\hbar D\partial^2_z\theta \approx 12\Gamma_\mathrm{sf}(E_g)\sin\theta\cos\theta.\label{eq4}
\Eeq
Under the assumption $d\ll L,\xi$, we can consider that the proximity angle in this region is constant, $\theta(z)\approx \theta(z=L)$ and  integrate Eq.~\eqref{eq4} arriving at
\Beq
&& \left.\partial_z\theta \right|_{z=L}= \left. - b^{-1}_g\sin\theta \cos\theta\right|_{z=L}, \label{eq6a} 
\Eeq
where $b_g=\hbar D/[12d\Gamma_{{\rm{sf}}}(E_g)]$ is the gate-dependent extrapolation length whose inverse determines the strength of the surface depairing. It is clear that the $E_g$ dependency of $b^{-1}_g$ follows from $\Gamma_{\rm{sf}}(E_g)$, i.e., for large $E_g$ we model it as $\xi/b_g=(E_g/E_{\rm{sc}})^4$, where $E_{\rm{sc}}$ is the scaling field at which $\xi/b_g=1$. The boundary condition at the free surface [$z=0$; see Fig.~\ref{schematic}(a)] follows from the current conservation law and simply reads $\partial_z\theta|_{z=0}=0$~\cite{KL}. Equation~\ref{eq6a} represents the central result of this work.  

\section{Results and discussion}\label{Sec:Results}

By solving Eq.~\eqref{eq3} supplemented by the derived boundary conditions [see Eq.~\eqref{eq6a}], we can describe the superconducting gap quenching and the supercurrent suppression caused by the gate-mediated surface depairing.

\begin{figure}[t!]
\includegraphics[width=0.95\linewidth]{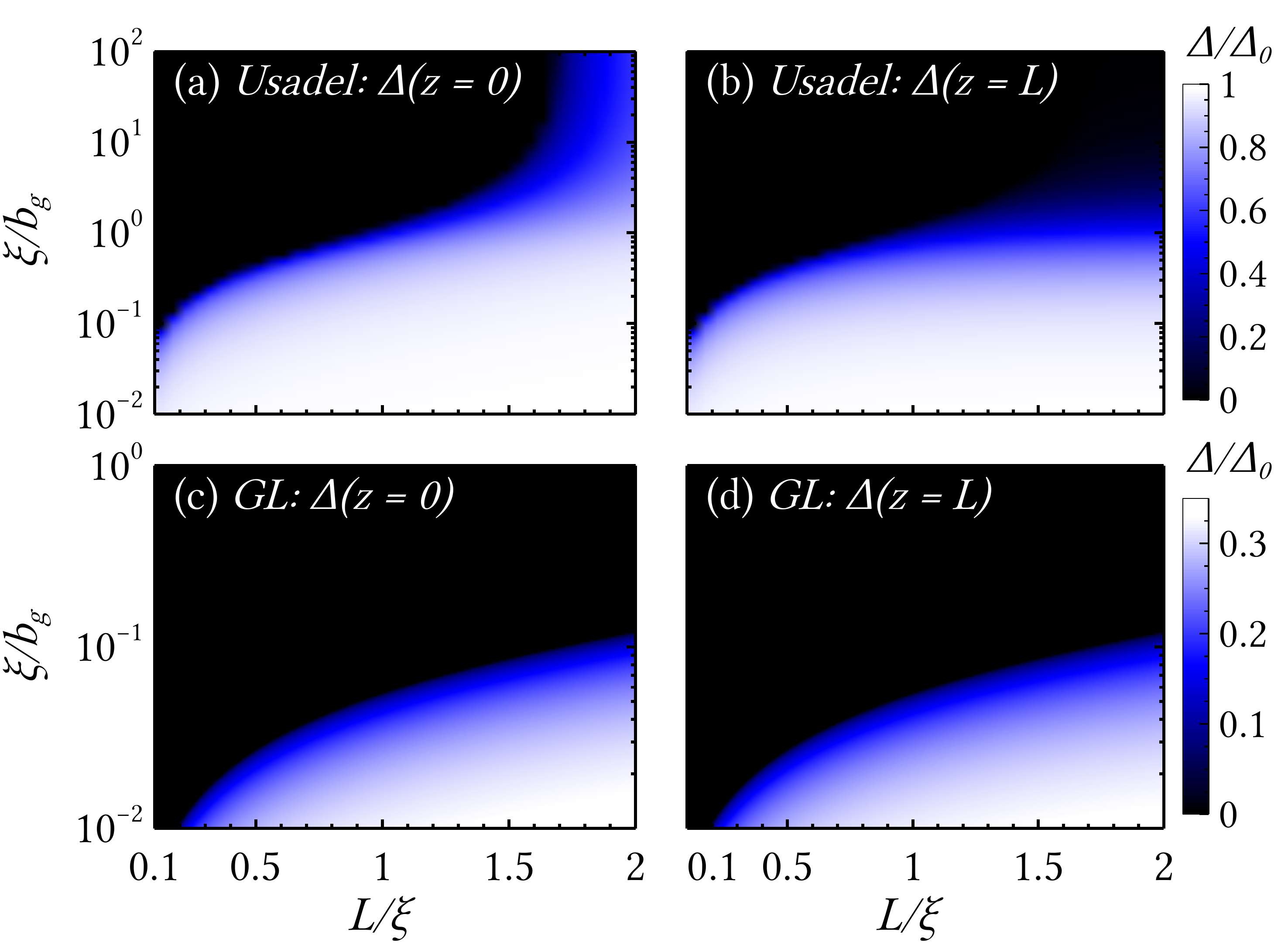}
\caption{\label{gapz} The gap function $\Delta(L,b^{-1}_g)$ at (a) $z=0$ (maximum value) and (b) $z=L$ (minimum value). Temperature is $T=0.1~T_c$.
Panels (c) and (d) respectively show the analytically obtained maximum and the minimum of the gap, calculated from the GL approach [see the text above Eq.~\eqref{eqn:GL}] at $T=0.95T_c$. In all panels, $q\xi=0.1$ and $B=0$.}
\end{figure}

Using Eq.~\eqref{eq6a} we present $\Delta (L,b^{-1}_g)$ in Fig.~\ref{gapz}(a,b). Since the gap is spatially dependent we illustrate the maximum [$z=0$; panel (a)] and the minimum values [$z=L$; panel (b)]. Apparently, in the case of thin bridges, $L\lesssim\xi$, a sufficiently high surface depairing can result in a complete quenching of the superconducting gap. The gap is almost spatially uniform across the bridge and diminishes monotonically with increasing $b^{-1}_g$. On the other hand, thicker bridges, $L>\xi$, feature a partial gap suppression. This is similar to an earlier theoretical finding \cite{Amoretti_GL}. Interestingly, for $L>\xi$ in the presence of sufficiently high $b^{-1}_g$ the gap can completely vanish only at the boundary while remaining non-zero inside the bridge. These observations demonstrate that the superconductivity of the bridge can be modulated by a direct field-controlled surface effect.

\begin{figure}
\includegraphics[width=1\linewidth]{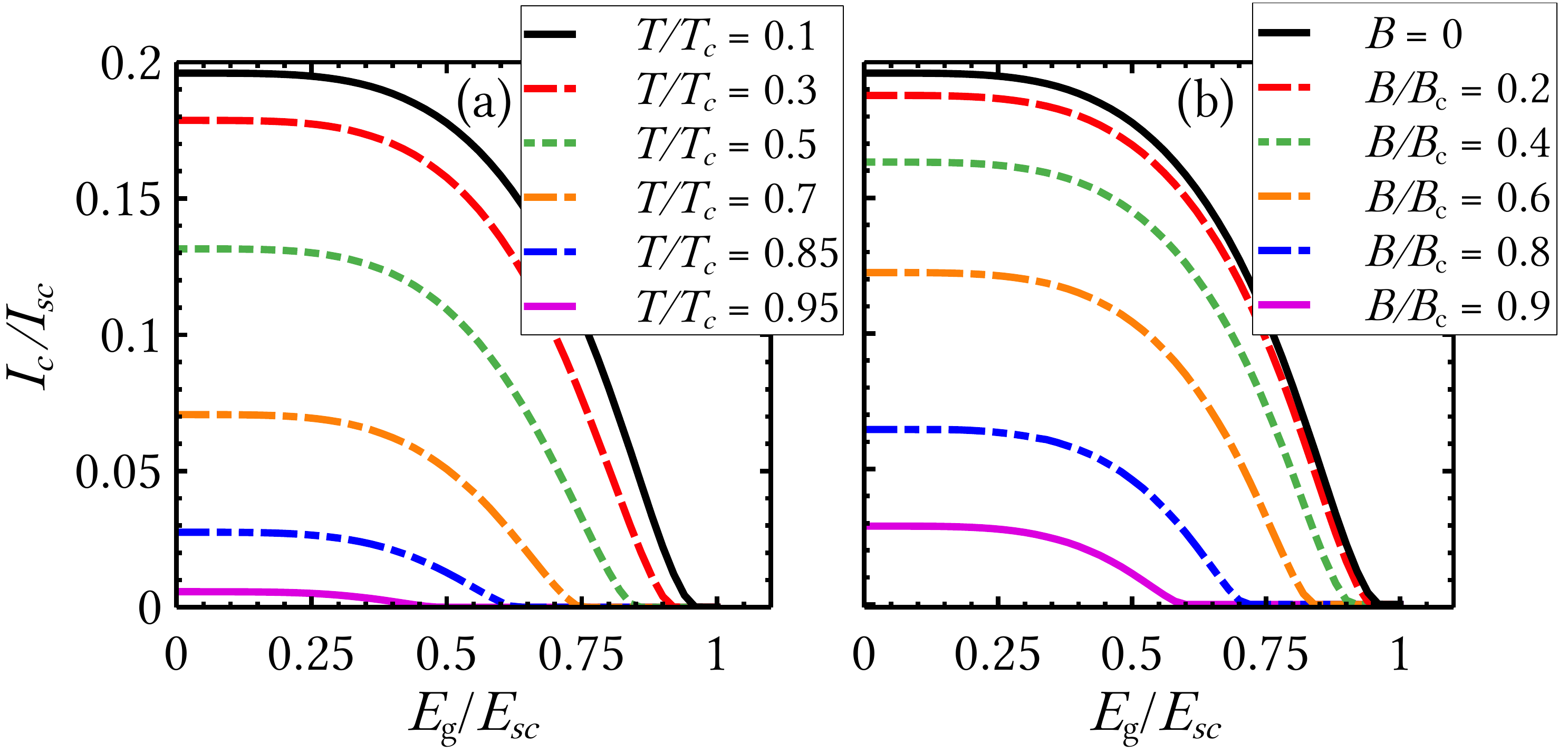}
\caption{\label{IcEg} The critical supercurrent $I_c$ vs. the gate field $E_g$ for (a) various temperatures $T$ and zero magnetic field $B=0$ and (b) various magnetic fields $B$ and temperature $T=0.1T_c$. The thickness of the bridge is $L=0.8\xi$.} 
\end{figure}

Following the Ginzburg-Landau (GL) approach close to $T_c$ for thin bridges, $L\lesssim\xi$, we obtain an approximate but analytic formula for the superconducting gap's maximum, $\Delta(z=0)=1.74\Delta_0(1-\mathcal{F}-T/T_c)^{1/2}$, and minimum, $\Delta(z=L)=(1+0.5L/b_g)^{-1}\Delta(z=0)$, where~(see Appendix~\ref{Appendix:GL})
\beq
\mathcal{F}= \frac{\pi}{4}\left(q^2\xi^2+2\frac{\Gamma_{\rm{orb}}}{k_BT_c}\right)+\frac{\pi\xi^2}{2(L^2+2Lb_g)}, \label{eqn:GL}
\eeq 
with the notation introduced before. From the expressions above, it is visible that increasing $b^{-1}_g$ causes the gap quenching. In Figs~\ref{gapz}(c) and \ref{gapz}(d) we show the analytically obtained maximum and minimum gaps, respectively, as a function of $L$ and $b_g^{-1}$. Note that the GL theory is, strictly speaking, valid only at temperatures close to $T_c$. Seemingly, our simplified model captures qualitatively the essential physics.
%%%%%%%%%%%%%
%%%%%%%%%%%%%%
Apparently, the quenching of $\Delta(z)$ due to surface depairing directly leads to $I_c$ suppression in the bridge. Figure~\ref{IcEg}(a) shows $I_c(E_g)$ for various $T$ and $B=0$. Note that we model the surface depairing by Eq.~\eqref{eq6a}, where for large gate field we consider $\xi/b_g=(E_g/E_{\rm{sc}})^4$, making $I_c$ symmetric with respect to $E_g$. 
Apparently, $I_c$ vanishes monotonically by increasing $E_g$. Higher temperature enhances the effect, i.e., the critical gate field is reduced. The latter is defined as the electric field for which the current completely vanishes. This effect is especially pronounced at high temperatures [see the blue dash-dotted and violet solid lines in Fig.\ref{IcEg}(a)], which is partly in disagreement with certain experiments \cite{Francesco2018,Boursprr,Francesco2020} that report the robustness of the critical field against temperatures. However, our model qualitatively captures $I_c(E_g)$ features at low temperatures, observed in experiments. The absence of temperature dependence in the experiments  
we briefly discuss below. In Fig.~\ref{IcEg}(b) we illustrate the role of $B$ on $I_c(E_g)$. Even small magnetic fields strongly enhance the supercurrent suppression and reduce the critical gate field.

Similarly to the temperature, the magnetic field affects the critical gate field stronger than that observed  experimentally, and the present theory cannot completely explain these deviations. However, accounting for $T$ and $B$ dependencies on the spin-flip processes occurring at the boundary may help to overcome these issues. We stress that the following discussion is purely phenomenological since there is still no microscopic mechanism for the $T$- and $B$-dependent spin-flip processes. 
To further extend the discussion about the above issues, one may consider $E_g$ as an effective electric field in the surface layer related to an actual external gate field, $E_{\rm{ext}}$, as $E_g=\chi(T,B)E_{\rm{ext}}$, where $\chi(T,B)$ is the relative permittivity in the surface layer. Consequently, $\chi$ can bring $T$- and $B$- dependencies in $\Gamma_{\rm{sf}}$ and hence in $\xi/b_g=\chi^{4}(T,B)(E_{\rm{ext}}/E_{\rm{sc}})^4$. The function $\chi$ can decrease as the system approaches the normal state from the superconducting one by increasing $T$ and $B$, e.g., $\chi(T_c,B=0)=\chi(T=0,B_c)=0$~\cite{Francesco2018}. Introducing $\chi$ we may achieve weaker $T$- and $B$- dependencies of the critical values of $E_{\rm{ext}}$ (see Appendix~\ref{Appendix:Phenomenology}). However, a detailed microscopic analysis of $\chi(T,B)$ is required for further understanding.

Finally, to provide some realistic values for the parameters of our model, let us consider an aluminum superconductor characterized by the critical temperature $T_c\approx 1.2$~K, the density of states $\nu\approx 2\times 10^{47}$ J$^{-1}$m$^{-3}$, and the lattice constant $a\approx 4~\AA$. By assuming the typical $D\approx 20$ cm$^2$/s, we end up with $\xi\approx 80$~nm. In experiments, typical values of the gate field are rather large, $E_g\sim 700$~MV/m and the Thomas-Fermi screening length is typically small, $d\sim 1~$nm. Approximating hopping energy as $h_{ij}\sim eE_ga$ and considering electron pair repulsion at a magnetic impurity site $u\approx1$ eV, we may estimate $J_{ij}=h^2_{ij}/2u$ and $J(\Theta)$ as $\sim J_{ij}a^3$. Taking all these parameters into account brings us to the estimated magnetic impurity concentration in the surface layer of $N_m\approx 2~\mathrm{nm}^{-3}~$nm that corresponds to the surface depairing strength $\xi/b_g\approx0.1$. As our discussion presented above suggests this value of $\xi/b_g$ suffices to significantly reduce $I_c$ in a thin bridge, e.g., $L\approx0.5\xi$.
Hence, according to our model's a surface magnetic impurity density $2\times 10^{14}$ cm$^{-2}$ would be enough to observe a significant gate-controlled supercurrent suppression. We note that the effective exchange correlation length larger than $a$ can cause sizable $J_{ij}$ for lower values of gate fields.  

\section{Conclusions}\label{Sec:Conclusions}

We conclude that if a superconducting bridge features dilute magnetic impurities on the surface, a large gate field can significantly enhance spin-flip scattering at the surface magnetic impurity centers.
Consequently, these magnetic impurities can lead to a significant enhancement of the surface depairing via gate-induced spin-flip scattering resulting in the GCS suppression. We find that for thin bridges, $L\lesssim \xi$, one can achieve the complete suppression beyond some strong gate field. Yet, thicker bridges, $L>\xi$, can only feature a partial suppression. These findings suggest that the GCS suppression could be a consequence of the surface depairing. Our model also captures the bipolar behavior with respect to the gate-field direction, one of the key features of the direct field effect \cite{Francesco2018, Francesco2020, Francesco2021,Boursprr}. In addition, we have analyzed the impact of temperature and a weak magnetic field on the critical supercurrent and the critical gate field. Our findings indicate that the supercurrent suppression could originate from a direct gate effect. The temperature and magnetic field dependence of our model reproduces qualitatively experimental data. Future experiments can test our predictions, e.g., by artificially modifying the superconducting surface with magnetic impurities and correlating their concentration with our model estimates.  

\acknowledgements 

We acknowledge support from the EU’s Horizon 2020 research and innovation program under Grant Agreement No. 964398 (SUPERGATE) and from Deutsche Forschungsgemeinschaft (DFG; German Research Foundation) via SFB 1432 (project No. 425217212).

SC and DN contributed equally to the work.

\appendix

\section{Ginzburg-Landau (GL) description}\label{Appendix:GL}

In Eq.~\eqref{eq3}, the Green's functions are parametrized by the proximity angle $\theta$, where $g=\cos \theta$ and $f=\sin\theta$. Near the superconducting critical temperature $T_c$, the anomalous Green function $f$ and the superconducting gap $\Delta(z)$ are small. Hence, we can consider $f=\sin\theta \approx \theta$ and the Usadel equation~\eqref{eq3} can be approximated as  
\Beq
&& \hbar D\partial^2_zf=2\omega_nf -2\Delta(z)g +2\Gamma_\mathrm{eff}(q)gf, \label{jan10b}
\Eeq
where $\Gamma_\mathrm{eff}(q)= \hbar D(q^2/2) +2\Gamma_{\rm{orb}}$ inside the bridge. Close to $T_c$ and for small $q$ and $B$, we can consider $g\approx g_0=\omega_n/\sqrt{\omega_n^2 +\Delta^2(z)}$. On the other hand, we can express the anomalous Green function as $f=f_0+f_1$, where $f_{0}$ is zeroth order in gradients and $f_1$ is the first correction which is quadratic in gradients, i.e., $f_1\sim \partial_z^2f_0$. By returning this expansion into Eq.~\eqref{jan10b}, we end up with the following relations:
\Beq
&& f_0=\frac{\Delta(z)}{\omega_n+\Gamma_\mathrm{eff}(q) ~ g_0}g_0,  \label{jan10c} \\
&& f_1=\frac{\hbar D}{2(\omega_n+\Gamma_\mathrm{eff}(q) ~ g_0)} \partial^2_zf_0. \label{jan10d}
\Eeq
Moreover, close to $T_c$ the gap is small, $\Delta\ll k_BT_c$, and the equations above can be further expanded
\Beq
&& f_0=\frac{\omega_n}{\tilde{\omega}_n}\left[\frac{\Delta(z)}{\omega_n}-\frac{1}{2}\left(\frac{\Delta(z)}{\omega_n}\right)^3 \right],  \label{jan10c} \\
&& f_1=\frac{\hbar D}{2\tilde{\omega}_n} \partial^2_zf_0, \label{jan10d}
\Eeq
where $\tilde{\omega}_n=\omega_n+\Gamma_\mathrm{eff}(q)$. Note that we aim at the expansion up to the third order in $\Delta$ and the second order in gradients.

By inserting Eqs.~\eqref{jan10c} and \eqref{jan10d} back into the self-consistency equation~\eqref{eq8}, we obtain
\Beq
\Delta(z)\ln (T/T_c) &=& 2\pi k_BT\sum_{n=0}^{N_D}\left[\frac{\Delta(z)}{\omega_n} \left(\frac{\omega_n}{\tilde{\omega}_n}-1\right) \right. \nonumber \\
&& \left. -\frac{1}{2}\frac{\omega_n}{\tilde{\omega}_n}\left(\frac{\Delta(z)}{\omega_n}\right)^3 + \frac{D}{2\tilde{\omega}_n} \partial^2_z\left(\frac{\Delta(z)}{\tilde{\omega}_n}\right) \right]. ~~~ \label{jan10e}
\Eeq
For small superconducting phase gradients, $q\xi\ll 1$, and weak magnetic fields, $\Gamma_\mathrm{orb}\ll k_BT_c$, the pair-breaking rate $\Gamma_\mathrm{eff}(q)$ is considerably smaller than Matsubara frequencies $\omega_n=(2n+1)\pi k_B T$ near $T_c$. Hence, we can take
\Beq
\frac{1}{\tilde{\omega}_n}\approx \frac{1}{{\omega}_n} - \frac{\Gamma_\mathrm{eff}}{\omega_n^2}.  \label{jan10f}
\Eeq 
Using Eqs.~\eqref{jan10e} and \eqref{jan10f}, along with the definition of the Riemann $\zeta$-function as $\zeta(p)=\sum_{n=1}^\infty (1/n^{p})$, we obtain the Ginzburg-Landau (GL) equation as follows:
\Beq
 \tilde{\xi}^2\partial^2_z\Delta(z) +\alpha\Delta(z)-\beta\Delta^3(z)=0, \label{jan10g} 
\Eeq 
where
\begin{widetext}
\Beq
&& \tilde{\xi}^2=\hbar D\left[\frac{3}{4\pi k_BT}\zeta(2) + \frac{15 ~\Gamma^2_\mathrm{eff}(q)}{16\pi^3k^3_BT^3}\zeta(4) -\frac{14~ \Gamma_\mathrm{eff}(q)}{8\pi^2k^2_BT^2}\zeta(3) \right]  \nonumber \\
&& ~~~~ \approx \hbar D\left[\frac{3}{4\pi k_BT_c}\zeta(2) + \frac{15~ \Gamma^2_\mathrm{eff}(q)}{16\pi^3k^3_BT_c^3}\zeta(4) -\frac{14 ~ \Gamma_\mathrm{eff}(q)}{8\pi^2k^2_BT_c^2}\zeta(3) \right]~~\text{close~to~}~T_c, \label{jan10h} \\
&& \alpha=-\frac{3~\Gamma_\mathrm{eff}(q)}{2\pi k_BT}\zeta(2) -\ln(T/T_c)  \approx \left(1-\frac{T}{T_c}\right) -\frac{3~\Gamma_\mathrm{eff}(q)}{2\pi k_BT_c}\zeta(2) ~~\text{close~to~} T_c, \label{jan10i} \\
&& \beta = \frac{7}{8\pi^2k^2_BT^2}\zeta(3) -\frac{15~\Gamma_\mathrm{eff}(q)}{16\pi^3k^3_BT^3}\zeta(4) \approx \beta_0 -\frac{15~\Gamma_\mathrm{eff}(q)}{16\pi^3k^3_BT_c^3}\zeta(4) ~~\text{close~to~}~T_c~\text{with}~~ \beta_0=\frac{7}{8\pi^2k^2_BT_c^2}\zeta(3). \label{jan10j}
\Eeq
\end{widetext}
In order to solve the differential equation in Eq.~\eqref{jan10g}, we need two boundary conditions. At the free surface, we have 
\Beq
\left.\partial_z\Delta(z)\right|_{z=0}=0, \label{jan10ja}
\Eeq
which is nothing but the current conservation law.
The other, gate-dependent, boundary condition can be obtained as follows:
\Beq
&& \left. \partial_zf(z,\omega_n)\right|_{z=L}=-\frac{1}{b_g}f(z=L,\omega_n), \label{jan10k} \\
\implies && \partial_z\sum_{n=0}^\infty f(z,\omega_n)\bigg|_{z=L}=-\frac{1}{b_g}\sum_{n=0}^\infty f(z=L,\omega_n), ~~~ \label{jan10l} \\
\implies &&  \partial_z \Delta(z)\big|_{z=L}=-\frac{1}{b_g}\Delta(z=L) \label{jan10m},
\Eeq
where $b_g$ was defined earlier [see Eq.~\eqref{eq6a}].
With Eqs.~\eqref{jan10g}-\eqref{jan10j} and the boundary conditions in Eqs.~\eqref{jan10ja} and \eqref{jan10m}, one can analyze gate-mediated superconductivity suppression in a superconducting nano-bridge close to $T_c$.

The GL description provided above is advantageous since it allows us to treat the problem analytically under certain assumptions. Namely, if the bridge is sufficiently thin, $L<\xi$, we can assume a weak spatial dependence of the order parameter $\Delta(z)$ and introduce the following ansatz:
\beq
\Delta(z)\approx \Delta(z=0)+\frac{\Delta_2}{2}z^2+\cdots, \label{jan19a}
\eeq
where the linear term in $z$ is zero due to the boundary condition at the free surface [see Eq.~\eqref{jan10m}]. Note that the second term is small compared to the leading one by a factor $\sim L^2/\xi^2$ and subsequent terms are even smaller. Using the boundary condition given by Eq.~\eqref{jan10m}, for thin bridges we obtain
\beq
\Delta_2=-\frac{2\Delta(z=0)}{2b_gL +L^2}, \label{jan19b}
\eeq
where $b_g$ was defined earlier.
Then by substituting Eq.~\eqref{jan19a} in the GL equation [see Eq.~\eqref{jan10g}] and using Eq.~\eqref{jan19b}, we obtain the following solution for the superconducting order parameter:
\Beq
\Delta(z) = \sqrt{\frac{\alpha}{\beta}-\frac{2\tilde{\xi}^2}{(2b_gL+L^2)\beta}}\left(1-\frac{1}{2b_gL+L^2}z^2\right). \label{jan19d}
\Eeq
This rather simple result gives a qualitative insight into the effect of the interface pair breaking. Namely, in the case of long junctions, $L\gg \tilde{\xi}$, we immediately notice that the solution reduces to the bulk solution, $\Delta(z)=\Delta_\infty = \sqrt{\alpha/\beta}$. The same holds if the pair-breaking rate is very weak, $b_g\gg\tilde{\xi}$. 

\section{Gate-mediated critical current suppression with temperature and magnetic field dependent spin-flip scattering: phenomenology}\label{Appendix:Phenomenology}

\begin{figure}[b!]
	\includegraphics[width=1\linewidth]{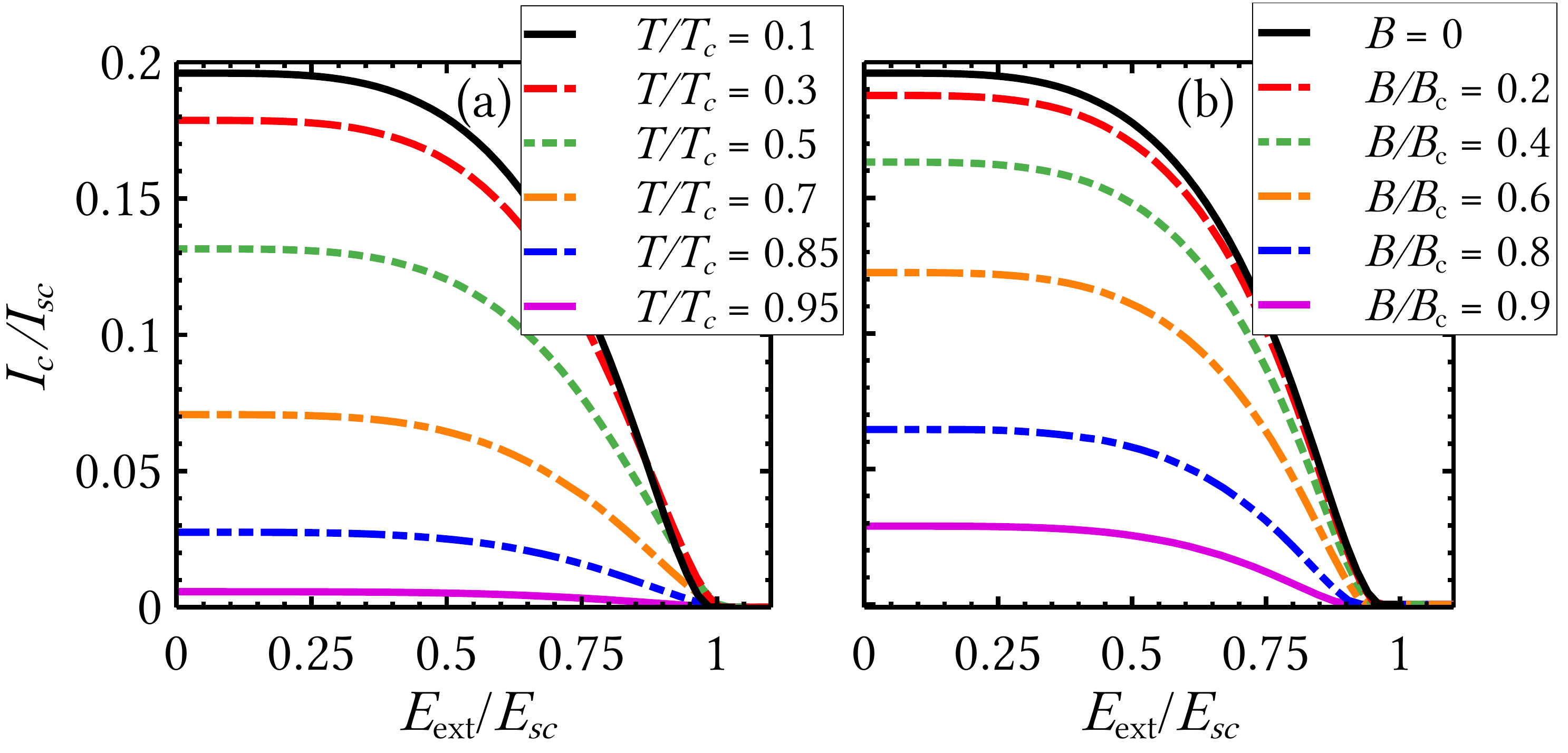}
	\caption{\label{IcEg2} Panel (a): The critical current $I_c$ as a function of the external gate field $E_{\rm{ext}}$ for various temperatures $T$, the bridge's thickness $L=0.8\xi$ and zero magnetic field, $B=0$. Panel (b): the same quantity for various magnetic fields $B$ and $T/T_c = 0.1$, other parameters are the same as in panel (a). Current is expressed in the unit of $I_{\rm{sc}}=2\pi k_BT_c/eR_N$.}
\end{figure}

To account for the full temperature dependency of the critical current suppression via the external gate field, we may think of $E_g$ as an effective electric field in the surface layer. The effective field $E_g$ affecting the superconducting bridge could be different from the actual external gate field, $E_{\rm{ext}}$~\cite{Francesco2018}. We can consider $E_g=\chi(T,B) E_{\rm{ext}}$, where the relative permittivity $\chi$ can depend on $T$ and $B$. Earlier scientific work demonstrates that the electric field screening effect increases as the system approaches from superconducting to normal state \cite{Francesco2018}. Therefore in the absence of $B$ the effective gate field in the surface layer would be maximum at $T=0$~K and it would be negligibly small in the  normal metal state ($T\geq T_c$). 
Phenomenologically the relative permittivity can be modeled as $\chi(T,B=0)=(1-T/T_c)^\eta$ with $\eta>0$. Apparently, $\chi(T=0,B=0)=1$ and $\chi(T=T_c,B=0)=0$ \cite{Francesco2018}.
Hence, in the absence of $B$, the surface depairing parameter $b_g^{-1}$ is temperature dependent having the form $\xi/b_g=(1-T/T_c)^{4\eta}(E_{\rm{ext}}/E_{\rm{sc}})^4$. For illustration, Fig.~\ref{IcEg2}(a) shows the the critical current $I_c$ vs. the external field $E_\mathrm{ext}$ for various temperatures $T$, the bridge's thickness $L=0.8\xi$, and $\eta=1/4$. Note that the critical external field is quite stable with respect to temperature. 

To account for the magnetic field dependency on the critical current, similarly, we can choose $\chi(T,B)$ for a fixed temperature. However, in this case, the situation is somewhat complicated. Namely, to provide stability with respect to $B$ we are supposed to introduce two fitting parameters, $\eta_1$ and $\eta_2$, i.e., $\chi(T,B)=\left[1-(B/B_c)^{\eta_1}\right]^{\eta_2}$. Figure~\ref{IcEg2}(b) shows $I_c$ vs. $E_{\rm{ext}}$ for various $B$, $T/T_c=0.1$, and $\eta_1=2$ and $\eta_2=1/4$.

\end{document}